\documentclass[a4paper,12pt,epsf,amssymb,citesort]{article}

\usepackage{tabularx}
\usepackage{array}
\usepackage{graphics}
\usepackage{graphicx}
\usepackage{epsfig}
\usepackage{amsmath}
\usepackage{amssymb}
\usepackage{cite}
\usepackage{verbatim}

\def\lsim{\raise0.3ex\hbox{$<$\kern-0.75em\raise-1.1ex\hbox{$\sim$}}}
\def\gsim{\raise0.3ex\hbox{$>$\kern-0.75em\raise-1.1ex\hbox{$\sim$}}}
\def\noi{\noindent}  \def\bea{\begin{eqnarray}}
\def\eea{\end{eqnarray}} \def\beq{\begin{equation}}
\def\eeq{\end{equation}} 
\def\beeq{\begin{eqnarray}} \def\eeeq{\end{eqnarray}} \def\R{ {\rm R
\kern -.31cm I \kern .15cm}} \def\C{ {\rm C \kern -.15cm \vrule
width.5pt \kern .12cm}} \def\Z{ {\rm Z \kern -.27cm \angle \kern
.02cm}} \def\N{ {\rm N \kern -.26cm \vrule width.4pt \kern .10cm}}
\def\1{{\rm 1\mskip-4.5mu l} }

\begin{document}

\begin{center} 

\par \vskip 10 truemm

{\Large \bf $B$ decays to radially excited $D$ mesons }
\par \vskip 2.5 truemm
{\Large \bf in Heavy Quark Effective Theory}

\par \vskip 8 truemm

 {\sc Damir Be\v{c}irevi\'c, Alain Le Yaouanc and Luis Oliver}

\par \vskip 8 truemm

{\sl Laboratoire de Physique Th\'eorique (B\^at. 210)} \footnote{Unit\'e Mixte de Recherche UMR 8627 - CNRS}
\vskip 1.6 truemm
{\sl Univ. Paris-Sud, Universit\'e de Paris-Saclay, 91405 Orsay, France.}

\end{center}

\par \vskip 2 truemm

\begin{abstract}

Semileptonic transitions $\overline{B} \to D^{(n)} \ell \overline{\nu}$, where $D^{(n)} (n \not = 0)$ is a radially excited meson, have recently attracted much attention as a way to understand some puzzles between theory and data. Following closely the formalism of Falk and Neubert for the elastic case, we study the $1/m_Q$ corrections to the heavy quark limit, in which the inelastic Isgur-Wise function vanishes at zero recoil, $\xi^{(n)}(1) = 0\ (n \not = 0)$. We find simple formulas that involve the derivative $\xi^{(n)'}(1)$, and we propose a number of ways of isolating this quantity in practice. We formulate also a generalization to the inelastic case of Luke's theorem. On the other hand, although some $1/m_Q$ HQET results are satisfied in the Bakamjian-Thomas relativistic quark model, we emphasize the problems concerning these corrections in this scheme.

\end{abstract}

\par \vskip 5 truemm

\noi {\small LPT-Orsay-17-21}

\par \vskip 3 truemm

\noi {\footnotesize Damir.Becirevic@th.u-psud.fr}\par
\noi  {\footnotesize Alain.Le-Yaouanc@th.u-psud.fr}\par
\noi  {\footnotesize Luis.Oliver@th.u-psud.fr}

\par \vskip 2 truemm

\section{Introduction}

This note is devoted to the discussion of $1/m_Q$ corrections in Heavy Quark Effective Theory (HQET) for the heavy quark current transitions between the ground state and radially excited states, e.g. $\overline{B} \to D^{(n)}\ (n \not= 0)$.\par
These transitions have been the object of some attention in recent years, in the view of solving some puzzles of semileptonic $B$ decays \cite{LIGETI,BBGLS}.\par
Rather detailed calculations of transition form factors to heavy meson radial excitations have been performed in Refs.~ \cite{GALKIN, FG} in the framework of a relativistic quark model. See also Ref.~\cite{SEGOVIA}.\par
The IW function $\xi^{(n)}(w)$ for the transitions $0 \to n$ satisfies the normalization condition
\beq
\label{7e}
\xi^{(n)}(1) = \delta_{n,0}\ ,
\eeq

\noi that expresses the normalization of the elastic IW function and the orthogonality between the initial and the final states for $n \not= 0$.\par

Our purpose here is to study in detail the subleading $1/m_Q$ corrections in HQET, following closely and in a straightforward way the paper by Falk and Neubert devoted to the elastic case \cite{FN}.

\vskip 0.5cm

\section{Form factors to radially excited states in HQET including $1/m_Q$ corrections}

\vskip 0.5cm

Besides the leading inelastic Isgur-Wise function $\xi^{(n)}(w)\ (n > 0)$, we consider the first order $1/m_Q$ corrections, that are of two types, corresponding to corrections to the heavy quark current, and to corrections to the leading order HQET Lagrangian. 

\subsection{$1/m_Q$ corrections of the Current type} 

 Consider first the leading and subleading corrections to the quark current
\beq
\label{8e}
\overline{c} \Gamma b \to \overline{h}_c \Gamma h_b + {1 \over 2m_c}\ \overline{h}_c \Gamma \gamma^\alpha \left( iD_\alpha h_b \right) + {1 \over 2m_b}\ \overline{\left( iD_\alpha h_c \right)} \gamma^\alpha \Gamma h_b \,,
\eeq

\noi where $\Gamma$ is a Dirac matrix. The matrix elements of the relevant operators read, in the matrix formalism of ref. \cite{FN}:
\beq
\label{9e}
< D^{(n)}(v')\mid\overline{h}_c \Gamma h_b \mid \overline{B}(v) >\ = - \xi^{(n)}(w) Tr[\overline{\mathcal{D}} \Gamma \mathcal{B}]\,,
\eeq
\beq
\label{10e}
< D^{(n)}(v')\mid\overline{h}_c \Gamma \gamma^\alpha \left( iD_\alpha h_b \right)\mid \overline{B}(v) >\ = - Tr[\xi^{(b)(n)}_\alpha(v,v')\overline{\mathcal{D}} \Gamma \gamma^\alpha \mathcal{B}]\,,
\eeq
\beq
\label{11e}
< D^{(n)}(v')\mid\overline{\left( iD_\alpha h_c \right)} \gamma^\alpha \Gamma h_b\mid \overline{B}(v) >\ = - Tr[\overline{\xi}^{(c)(n)}_\alpha(v',v)\overline{\mathcal{D}} \gamma^\alpha \Gamma \mathcal{B}]
\,, \eeq

\noi where $\overline{A} = \gamma^0 A^\dagger \gamma^0$ is the Dirac conjugate matrix, $\mathcal{B} = P_+ (-\gamma_5), \mathcal{D} = P'_+ (-\gamma_5)$ are given in terms of the projectors $P_+ = {1+{/\hskip - 2 truemm v} \over 2}, P'_+ = {1+{/\hskip - 2 truemm v^\prime} \over 2}$, where $v$ and $v'$ are the initial and final four-velocities, $D^{(n)}(v')$ denotes the pseudoscalar $D^{(n)}$ or vector $D^{(n)*}$ final states, and the functions $\xi^{(Q)(n)}_\alpha(v,v')$ ($Q = b, c$) read, in all generality:
\beq
\label{12e}
\xi^{(Q)(n)}_\alpha(v,v') = \xi_+^{(Q)(n)}(w)(v+v')_\alpha + \xi_-^{(Q)(n)}(w)(v-v')_\alpha - \xi_3^{(Q)(n)}(w)\gamma_\alpha \ .
\eeq

Following \cite{FN}, from the equations of motion and the matrix element of the divergence of the current one obtains
\beq
\label{20e}
\xi_-^{(b)(n)}(w) = {\overline{\Lambda}w-\overline{\Lambda}^{(n)} \over 2(w-1)}\ \xi^{(n)}(w) \ ,
\eeq
\beq
\label{21e}
\xi_-^{(c)(n)}(w) = - {\overline{\Lambda}-\overline{\Lambda}^{(n)}w \over 2(w-1)}\ \xi^{(n)}(w)\ .
\eeq

Taking the limit $\overline{\Lambda}^{(n)} \to \overline{\Lambda}$ while keeping $w$ fixed, one finds the results of the elastic case \cite{FN}, valid for all $w$, namely: 
\beq
\label{22e}
\xi_-^{(b)}(w) = \xi_-^{(c)}(w) = {\overline{\Lambda} \over 2}\ \xi(w)\ ,
\eeq

\noi where we have omitted the superindex $(0)$ corresponding to the ground state. 

In the inelastic case, taking the limit of relations (\ref{20e},\ref{21e}) for $w \to 1$, one obtains
\beq
\label{24e}
\xi_-^{(b)(n)}(1) = - \xi_-^{(c)(n)}(1) = - {\Delta E^{(n)} \over 2}\ \xi^{(n)'}(1),  \qquad \qquad (n > 0).
\eeq

\subsection{$1/m_Q$ corrections of the Lagrangian type} 

There are two types of $1/m_Q$ Lagrangian perturbations, depending on the insertion of the Lagrangian in the initial $b$ quark or in the final $c$ quark. Both types of perturbations, even in the case of the $b$ insertion, depend on the radial quantum number $n$ because the final state is radially excited :

$$<D(v') |i \int dxT [J^{cb} (0), {\cal L}_v^{(b)}(x)]|B(v)>\ = $$
\bea
\label{25e}
{1 \over 2m_b} \left \{ - A^{(b)(n)}(w) Tr \left [\overline{D}(v') \Gamma B(v)\right ] + {1 \over 2} Tr\left  [ A_{\alpha\beta}^{(b)(n)}(v,v') \overline{D}(v') \Gamma P_+ i \sigma^{\alpha\beta} B(v)\right ] \right \},
\eea

\vskip 0.3 truecm

$$<D(v') |i \int dxT [J^{cb} (0), {\cal L}_{v'}^{(c)}(x)]|B(v)>\ =$$
\bea
\label{26e}
{1 \over 2m_c} \left \{ - A^{(c)(n)}(w) Tr \left [\overline{D}(v') \overline{\Gamma} B(v)\right ] - {1 \over 2} Tr\left  [ A_{\alpha\beta}^{(c)(n)}(v',v) \overline{D}(v') i \sigma^{\alpha\beta} P'_+ \overline{\Gamma} B(v)\right ] \right \},
\eea

\noi with
\bea
\label{27e}
A_{\alpha \beta}^{(b)(n)}(v,v') = A_2^{(b)(n)}(w) \left ( v'_{\alpha} \gamma_{\beta} - v'_{\beta} \gamma_{\alpha}\right ) + A_3^{(b)(n)}(w) i \sigma_{\alpha \beta} \ ,
\eea
\bea
\label{28e}
\overline{A}_{\alpha \beta}^{(c)(n)}(v',v) = A_2^{(c)(n)}(w) \left ( v_{\alpha} \gamma_{\beta} - v_{\beta} \gamma_{\alpha}\right ) + A_3^{(c)(n)}(w) i \sigma_{\alpha \beta}\ ,
\eea

\noi where $\overline{A} = \gamma^0 A^+\gamma^0$ denotes the Dirac
conjugate matrix. The current $J^{cb}(0)$ denotes
\beq
\label{29e}
J^{cb} = \overline{h}_{v'}^{(c)} \Gamma h_v^{(b)}\ ,
\eeq

\noi where $\Gamma$ is any Dirac matrix, and ${\cal L}_v^{(Q)}(x)$ is given by
\beq
\label{30e}
{\cal L}_v^{(Q)} = {1 \over 2 m_Q} \left [ O_{kin, v}^{(Q)} + O_{mag, v}^{(Q)}\right ]
\eeq

\noi with
\beq
\label{31e}
O_{kin, v}^{(Q)} = \overline{h}_v^{(Q)}(iD)^2 h_v^{(Q)} , \qquad O_{mag, v}^{(Q)} = {g_s \over 2} \overline{h}_v^{(Q)} \sigma_{\alpha \beta} G^{\alpha\beta} h_v^{(Q)} \ .
\eeq

\subsection{Radially excited form factors at order $1/m_Q$}

Generalizing the notation of Ref.~\cite{FN}, considering now radial excitations in the final state $\overline{B} \to D^{(n)(*)} \ell \nu$ ($n = 0$ is the quasi-elastic case), the different form factors read, up to first order in $1/m_Q$ : 
\beq
\label{32e}
h_+^{(n)} = \xi^{(n)} + \epsilon_c L_1^{(c)(n)} + \epsilon_b L_1^{(b)(n)},
\eeq
\beq
\label{33e}
h_-^{(n)} = \epsilon_c L_4^{(c)(n)} - \epsilon_b L_4^{(b)(n)},
\eeq
\beq
\label{34e}
h_V^{(n)} = \xi^{(n)} + \epsilon_c \left(L_2^{(c)(n)} - L_5^{(c)(n)}\right) + \epsilon_b \left(L_1^{(b)(n)} - L_4^{(b)(n)}\right) ,
\eeq
\beq
\label{35e}
h_{A_1}^{(n)} = \xi^{(n)} + \epsilon_c \left( L_2^{(c)(n)} - {w-1 \over w+1}\ L_5^{(c)(n)} \right) + \epsilon_b \left( L_1^{(b)(n)} - {w-1 \over w+1}\ L_4^{(b)(n)} \right) ,
\eeq
\beq
\label{36e}
h_{A_2}^{(n)} = \epsilon_c \left( L_3^{(c)(n)} + L_6^{(c)(n)} \right),
\eeq
\beq
\label{37e}
h_{A_3}^{(n)} = \xi^{(n)} + \epsilon_c \left( L_2^{(c)(n)} - L_3^{(c)(n)} - L_5^{(c)(n)} + L_6^{(c)(n)} \right) + \epsilon_b \left( L_1^{(b)(n)} - L_4^{(b)(n)} \right) ,
\eeq

\noi where $\xi^{(n)}(w)$ is the Isgur-Wise (IW) function for the transition between the ground state and the radially excited states $0 \to n$, $L_i^{(b)(n)}(w) (i = 1, 4)$ are elastic subleading form factors at order $\epsilon_b = 1/{2 m_b}$, and $L_i^{(c)(n)}(w) (i = 1,... , 6)$ are subleading form factors at order $\epsilon_c = 1/{2 m_c}$.\par

In terms of the functions defined above, one finds for the different functions $L_i^{(b)(n)}(w)\ (i = 1,... , 6)$:
$$L_1^{(b)(n)} = A_1^{(b)(n)} + 2(w-1)A_2^{(b)(n)} + 3A_3^{(b)(n)},$$
$$L_2^{(b)(n)} = A_1^{(b)(n)} - A_3^{(b)(n)},$$
\beq
\label{37e}
L_3^{(b)(n)} = -2A_2^{(b)(n)},
\eeq
$$L_4^{(b)(n)} = -  {\overline{\Lambda}w-\overline{\Lambda}^{(n)} \over w-1}\ \xi^{(n)}(w) + 2\xi_3^{(b)(n)},$$
$$L_5^{(b)(n)} = -  {\overline{\Lambda}w-\overline{\Lambda}^{(n)} \over w-1}\ \xi^{(n)}(w),$$
$$L_6^{(b)(n)} = -  {2 \over w+1} \left({\overline{\Lambda}w-\overline{\Lambda}^{(n)} \over w-1}\ \xi^{(n)}(w) + \xi_3^{(b)(n)}\right),$$

\noi and for the functions $L_i^{(c)(n)}(w)\ (i = 1,... , 6)$ :
$$L_1^{(c)(n)} = A_1^{(c)(n)} + 2(w-1)A_2^{(c)(n)} + 3A_3^{(c)(n)},$$
$$L_2^{(c)(n)} = A_1^{(c)(n)} - A_3^{(c)(n)},$$
\beq
\label{38e}
L_3^{(c)(n)} = -2A_2^{(c)(n)},
\eeq
$$L_4^{(c)(n)} = {\overline{\Lambda}w-\overline{\Lambda}^{(n)} \over w-1}\ \xi^{(n)}(w) + 2\xi_3^{(c)(n)},$$
$$L_5^{(c)(n)} = {\overline{\Lambda}-\overline{\Lambda}^{(n)}w \over w-1}\ \xi^{(n)}(w),$$
$$L_6^{(c)(n)} = -  {2 \over w+1} \left(- {\overline{\Lambda}w-\overline{\Lambda}^{(n)} \over w-1}\ \xi^{(n)}(w) + \xi_3^{(c)(n)}\right).$$

\subsection{Constraints on $1/m_Q$ radially excited form factors}

At order $1/m_Q$, in the transitions $0 \to n$ with $n \not= 0$, we have many more form factors than in the elastic case. However, some constraints still exist on these form factors.\par 

For the Current type form factors, we must emphasize the important corrections to heavy quark limit quantities: 
$$L_5^{(b)(n)}(w) = -  {\overline{\Lambda}w-\overline{\Lambda}^{(n)} \over w-1}\ \xi^{(n)}(w),$$
\beq
\label{38e}
L_5^{(c)(n)}(w) = {\overline{\Lambda}-\overline{\Lambda}^{(n)}w \over w-1}\ \xi^{(n)}(w),
\eeq

\noi and the constraints that follow from the equations (\ref{37e},\ref{38e}) by eliminating the unknown functions $\xi_3^{(b)(n)}(w), \xi_3^{(c)(n)}(w)$, namely: 
$$L_4^{(b)(n)}(w)+(w+1)L_6^{(b)(n)}(w) = 3L_5^{(b)(n)}(w),$$
\beq
\label{39e}
L_4^{(c)(n)}(w)+(w+1)L_6^{(c)(n)}(w) = 3L_5^{(c)(n)}(w).
\eeq

For the form factors of the Lagrangian type we do not have Luke's theorem \cite{LUKE}, but weaker generalizations of it.\par

Let us consider the transition between charmed pseudoscalars $D \to D^{(n)}$ through the elastic current $\overline{c} \gamma^0 c$ at zero recoil $w = 1$. From (\ref{32e}) one has
\beq
\label{40e}
h_+^{(n)}(1) = \xi^{(n)}(1) + \epsilon_c \left( L_1^{(c)(n)}(1) +  L_1^{(b)(n)}(1) \right) .
\eeq

\noi Recall that the different superscripts $(b)$ and $(c)$ mean simply that the initial and final states are different although the functions $L_1^{(c)(n)}, L_1^{(b)(n)}$ are independent of the heavy quark masses.\par

Since $\overline{c} \gamma^0 c$ is the $c$ quark number operator and one has the normalization (\ref{7e}), relation (\ref{39e}) reads
\beq
\label{41e}
\delta_{n,0} = \delta_{n,0} + \epsilon_c \left( L_1^{(c)(n)}(1) +  L_1^{(b)(n)}(1) \right).
\eeq

\noi It follows therefore 
\beq
\label{42e}
L_1^{(c)(n)}(1) +  L_1^{(b)(n)}(1) = 0\ .
\eeq

\noi For the elastic transition $n = 0 \to n = 0$ one has $L_1^{(c)(0)}(1) =  L_1^{(b)(0)}(1) = L_1(1)$, and one obtains
\beq
\label{43e}
L_1(1) = 0\ .
\eeq

Following \cite{FN}, considering likewise the transition between charmed vector mesons $D^* \to D^{(n)*}$ through the elastic current $\overline{c} \gamma^0 c$ at zero recoil $w = 1$, one has
\beq
\label{44e}
h_1^{(n)}(1) = \xi^{(n)}(1) + \epsilon_c \left( L_2^{(c)(n)}(1) +  L_2^{(b)(n)}(1) \right) ,
\eeq

\noi which implies
\beq
\label{45e}
\delta_{n,0} = \delta_{n,0} + \epsilon_c \left( L_2^{(c)(n)}(1) +  L_2^{(b)(n)}(1) \right) ,
\eeq

\noi i.e., 
\beq
\label{46e}
L_2^{(c)(n)}(1) +  L_2^{(b)(n)}(1) = 0 .
\eeq

\noi For the elastic transition $0 \to 0$ one has $L_2^{(c)(0)}(1) =  L_2^{(b)(0)}(1) = L_2(1)$, and therefore
\beq
\label{47e}
L_2(1) = 0 .
\eeq

The equations (\ref{43e},\ref{47e}) follow from Luke's theorem \cite{LUKE}, while relations (\ref{42e}) and (\ref{46e}) are the generalizations to radially excited inelastic transitions.

\subsubsection{Summary of the constraints on $1/m_Q$ radially excited form factors}

To summarize, in all generality one has the following constraints on the $1/m_Q$ corrections.\par

For the Current pertubations we have found :
$$L_5^{(b)(n)}(w) = -  {\overline{\Lambda}w-\overline{\Lambda}^{(n)} \over w-1}\ \xi^{(n)}(w) ,$$
\beq
\label{48e}
L_5^{(c)(n)}(w) = {\overline{\Lambda}-\overline{\Lambda}^{(n)}w \over w-1}\ \xi^{(n)}(w) .
\eeq

\noi The relations (\ref{48e}) imply, at zero recoil, for $n > 0$ :
\beq
\label{49e}
L_5^{(b)(n)}(1) = - L_5^{(c)(n)}(1) = \Delta E^{(n)} \xi^{(n)'}(1) \qquad \qquad (\Delta E^{(n)} = \overline{\Lambda}^{(n)} - \overline{\Lambda}) .
\eeq

On the other hand, we have found the linear relations
$$L_4^{(b)(n)}(w)+(w+1)L_6^{(b)(n)}(w) = 3L_5^{(b)(n)}(w) , $$
\beq
\label{50e}
L_4^{(c)(n)}(w)+(w+1)L_6^{(c)(n)}(w) = 3L_5^{(c)(n)}(w),
\eeq

\noi where the r.h.s. are given in terms of the functions (\ref{48e}). These relations imply at zero recoil :
$$L_4^{(b)(n)}(1)+2L_6^{(b)(n)}(1) = -\left(L_4^{(c)(n)}(1)+2L_6^{(c)(n)}(1)\right) ,$$
\beq
\label{51e}
= 3 \Delta E^{(n)}\ \xi^{(n)'}(1) \qquad (n > 0).
\eeq

For the Lagrangian perturbations we have found the following generalizations of Luke's theorem :
\beq
\label{52e}
L_1^{(c)(n)}(1) +  L_1^{(b)(n)}(1) = 0,
\eeq
\beq
\label{53e}
L_2^{(c)(n)}(1) +  L_2^{(b)(n)}(1) = 0.
\eeq

For completeness we write here the corresponding relations in the elastic case :
\beq
\label{54e}
L_5(w) = -\overline{\Lambda}\xi(w),
\eeq
\beq
\label{55e}
L_4(w)+(w+1)L_6(w) = 3L_5(w),
\eeq
\beq
\label{56e}
L_1(1) = L_2(1) = 0 .
\eeq

A remark is in order here. The inelastic relation (\ref{49e}) is not symmetric in the exchange $b \leftrightarrow c$ because the radial excitation occurs in the final charmed state. This gives an opposite sign between $L_5^{(b)(n)}(1)$ and $L_5^{(c)(n)}(1)$ for $n > 0$. This differs from the elastic case, for which one obtains (\ref{54e}) with $L_5^{(b)(0)}(w) = L_5^{(b)(0)}(w) = L_5(w)$ by taking the limit $\overline{\Lambda}^{(n)}  \to \overline{\Lambda}^{(0)} = \overline{\Lambda}$ at $w$ fixed.\par

Let us now compare the number of independent functions in the elastic $\overline{B} \to D^{(*)}$ and the inelastic cases $\overline{B} \to D^{(*)(n)} (n > 0)$ .\par
In the elastic case we have the leading IW function $\xi(w)$, the three Lagrangian perturbations $L_i(w) (i = 1,2,3)$ and the three Current perturbations $L_i(w) (i=4,5,6)$, that are reduced to one independent function, modulo the parameter $\overline{\Lambda}$ because of relations (\ref{54e},\ref{55e}). This leaves the IW function, four subleading functions (say $L_1(w), L_2(w), L_3(w), L_4(w)$) and the parameter $\overline{\Lambda}$. Because of Luke's theorem (\ref{56e}), at subleading order one is left at zero recoil with the three parameters $L_3(1), L_4(1), \overline{\Lambda}$.\par

In the inelastic case we have a priori the inelastic IW function $\xi^{(n)}(w)$ and the twelve subleading functions $L_i^{(c)(n)}(w), L_i^{(b)(n)}(w)  (i = 1,...,6)$. The four relations concerning Current perturbations (\ref{48e}-\ref{50e}) reduce the functions $L_i^{(c)(n)}(w), L_i^{(b)(n)}(w)  (i = 4,5,6)$ to two, say $L_4^{(c)(n)}(w), L_4^{(b)(n)}(w)$, modulo the parameters $\overline{\Lambda},\overline{\Lambda}^{(n)}$. Summarizing, we have therefore the inelastic IW function $\xi^{(n)}(w)$ and six subleading functions, $L_i^{(c)(n)}(w), L_i^{(b)(n)}(w)  (i = 1,2,3,4)$. At zero recoil we have the generalization of Luke's theorem giving the relations between Lagrangian perturbations (\ref{52e},\ref{53e}). One is left at zero recoil with the free parameters, $L_1^{(c)(n)}(1), L_2^{(c)(n)}(1), L_3^{(b)(n)}(1), L_3^{(c)(n)}(1)$, $L_4^{(b)(n)}(1), L_4^{(c)(n)}(1)$ together with $\overline{\Lambda},\overline{\Lambda}^{(n)}$ and the derivative of the inelastic IW function at zero recoil $\xi^{(n)'}(1)$, giving a total of 9 independent free parameters.

\section{Suggestions for the determination of the derivative at zero recoil  $\xi^{(n)'}(1)\ (n \not= 0)$}

The IW function for the inelastic transition $\xi^{(n)}(w)\ (n \not= 0)$ vanishes at zero recoil, eq. (\ref{7e}). Therefore, it would be very interesting to have information on the derivative at $w = 1$, at least for the first radial excitation. We have a number of suggestions for this purpose.\par

\vskip 4 truemm

(1) A first remark is the following. Using the second relation (\ref{49e}) 
\beq
\label{57-1e}
 L_5^{(c)(n)}(1) = -\Delta E^{(n)} \xi^{(n)'}(1) \qquad \qquad (n \not= 0),
\eeq

\noi and (\ref{34e},\ref{35e}) one has
\beq
\label{57-2e}
h_{V}^{(n)}(1) - h_{A_1}^{(n)}(1) = - \epsilon_c\ L_5^{(c)}(1) + O(\epsilon_b) = \epsilon_c\ \Delta E^{(n)} \xi^{(n)'}(1) + O(\epsilon_b) .
\eeq

\noi Therefore, up to $1/m_b$ corrections, this difference of form factors is proportional to $\Delta E^{(n)} \xi^{(n)'}(1)$, and could therefore give information on the derivative of the IW function at zero recoil.\par
Although difficult, one could in principle also envisage a lattice calculation of the difference of form factors at zero recoil (\ref{57-2e}) in the heavy quark limit for the $B$ meson ($m_b \to \infty$), and thus isolate the desired quantity.\par

\vskip 4 truemm

(2) Using the same argument, one could as well consider the ratio of the form factors, that gives
\beq
\label{57-2bise}
{h_{V}^{(n)}(1) \over h_{A_1}^{(n)}(1)} = - \epsilon_c\ L_5^{(c)}(1) + O(\epsilon_b) + O(\epsilon_c^2) = \epsilon_c\ \Delta E^{(n)} \xi^{(n)'}(1) + O(\epsilon_b) + O(\epsilon_c^2)  ,
\eeq

\noi that could also provide $\xi^{(n)'}(1)$ up to higher corrections. For example, varying the masses $m_b, m_c$ on the lattice, one could isolate this derivative.

\vskip 4 truemm

(3) Another possibility would be to use the identities (\ref{9e}-\ref{11e}), the definition (\ref{12e}) and the results (\ref{20e},\ref{21e}). 
After some algebra, one finds for $n \geq 0$ :
\beq
\label{57-3e}
< D^{(n)}(v')\mid\overline{h}_c i\overleftarrow {D}_\alpha (v-v')^\alpha h_b\mid \overline{B}(v) >\ = (w+1) (\overline{\Lambda}-\overline{\Lambda}^{(n)}w) \xi^{(n)}(w),
\eeq
\beq
\label{57-4e}
< D^{(n)}(v')\mid\overline{h}_c iD_\alpha (v-v')^\alpha h_b\mid \overline{B}(v) >\ = (w+1) (\overline{\Lambda}w-\overline{\Lambda}^{(n)}) \xi^{(n)}(w),
\eeq

\noi that gives, for the ground state $n = 0$ :
\beq
\label{57-5e}
\lim_{w \to 1} {< D(v')\mid\overline{h}_c i\overleftarrow {D}_\alpha (v-v')^\alpha h_b\mid \overline{B}(v) > \over 2(w-1)}\ = - \Lambda ,
\eeq
\beq
\label{57-6e}
\lim_{w \to 1} {< D(v')\mid\overline{h}_c iD_\alpha (v-v')^\alpha h_b\mid \overline{B}(v) > \over 2(w-1)}\ = \Lambda ,
\eeq

\noi while for $n > 0$ one finds
\beq
\label{57-7e}
\lim_{w \to 1} {< D^{(n)}(v')\mid\overline{h}_c i\overleftarrow {D}_\alpha (v-v')^\alpha h_b\mid \overline{B}(v) > \over 2(w-1)}\ = - \Delta E^{(n)} \xi^{(n)'}(1) ,
\eeq
\beq
\label{57-8e}
\lim_{w \to 1} {< D^{(n)}(v')\mid\overline{h}_c iD_\alpha (v-v')^\alpha h_b\mid \overline{B}(v) > \over 2(w-1)}\ = - \Delta E^{(n)} \xi^{(n)'}(1) .
\eeq

Although difficult, a consideration of these matrix elements on the lattice could in principle help isolating the interesting quantity $\Delta E^{(n)} \xi^{(n)'}(1)$ for radial excitations.\par

\vskip 4 truemm

(4) There is also a straightforward systematic method, namely to consider on the lattice a form factor like $h_+^{(n)}(w)$ (\ref{32e}) and vary $1/m_c$, $1/m_b$ to have access to the inelastic form factor, following the general argument proposed in ref. \cite{HEIN}. In this work, in the static limit for the $b$ quark, $h_+^{(n)}(w)$ has been clearly isolated for the first radial excitation, providing some information on this form factor. Of course, to get the derivative $\xi^{(n)'}(1)$ one would need to consider several values of $w$ near the zero recoil point $w = 1$.

\section{Subleading $1/m_Q$ corrections for radial excitations in the BT quark model}

As we have repeated on several occasions, the Bakamjian-Thomas (BT) relativistic quark model provides a covariant description of hadron wave functions in motion.
On the other hand, the model provides also covariant current matrix elements {\it in the heavy quark limit}, that satisfy Isgur-Wise scaling and heavy quark sum rules of the Bjorken-Uraltsev type (see for example the recent paper \cite{LEYAOUANC} and references therein) .\par 
However, at subleading order in $1/m_Q$, the model is no longer covariant. We have examined this problem in detail for the elastic transitions ${1 \over 2}^- \to {1 \over 2}^-$ and also for inelastic ones to $L = 1$ orbital excitations ${1 \over 2}^- \to {1 \over 2}^+$, ${1 \over 2}^- \to {3 \over 2}^+$ \cite{DONG}.\par 
Although the model is not covariant in general, we have found a number of encouraging results for the $1/m_Q$ subleading form factors in the elastic case, ${1 \over 2}^- \to {1 \over 2}^-$. These include Luke's theorem and several other interesting relations at zero recoil like $L_5(1) = -\overline{\Lambda}$ \cite{DONG}.\par
However, we have found that serious difficulties appear for the inelastic transitions ${1 \over 2}^- \to {1 \over 2}^+$, ${1 \over 2}^- \to {3 \over 2}^+$. The most relevant ones concern subleading form factors that, at zero recoil, can be expressed in terms of level spacings and leading IW functions \cite{LEIBOVICH}. These relations, that are consequences of the QCD dynamics, are not fulfilled by the BT quark model scheme, as discussed in detail in \cite{DONG}.\par
For the transitions to radial excitations ${1 \over 2}^- (n = 0) \to {1 \over 2}^- (n \not = 0)$ we find a similar situation as for the orbital excitations.  Unlike what we have done in great detail in \cite{DONG} for the latter, we do not pretend here to compute the $w$ dependence of the form factors including the next-to-leading $1/m_Q$ corrections. We will restrict ourselves to some examples of subleading form factors at zero recoil that illustrate the situation in this case.\par
Let us consider the counterparts in the BT model of some relations for the subleading form factors within HQET, namely (\ref{42e},\ref{46e}), that are generalizations to the inelastic radially excited case of Luke's theorem, and (\ref{49e}), that involve level spacings and leading order IW functions.

To isolate $L_1^{(c)(n)}(1) +  L_1^{(b)(n)}(1)$ from (\ref{42e}) we need to consider the form factor to transitions to radial excitations $h_+^{(n)}(w)$, that can be easily read from the generalization of (B1) of Ref.~\cite{FN}:
\beq
\label{57e}
h_+^{(n)} = \xi^{(n)} + \epsilon_c L_1^{(c)(n)} + \epsilon_b L_1^{(b)(n)} + O(1/m_Q^2) .
\eeq

Computing the form factor $h_+^{(n)}(w)$ in the BT model, and making formally $\epsilon_c = \epsilon_b$ we find at zero recoil the relation (\ref{42e})
\beq
\label{58e}
L_1^{(c)(n)}(1) +  L_1^{(b)(n)}(1) = 0 .
\eeq

Considering now the transition $\overline{B}^* \to D^*$, with the vector current and longitudinal polarizations, one has similarly
\beq
\label{59e}
h_1^{(n)} = \xi^{(n)} + \epsilon_c L_2^{(c)(n)} + \epsilon_b L_2^{(b)(n)} + O(1/m_Q^2) ,
\eeq

Computing the form factor $h_1^{(n)}(w)$ in the BT model, and making $\epsilon_c = \epsilon_b$ we find at zero recoil the relation (\ref{46e})
\beq
\label{60e}
L_2^{(c)(n)}(1) +  L_2^{(b)(n)}(1) = 0 .
\eeq

Therefore, the generalizations to the inelastic case of Luke's theorem are fulfilled in the BT model.\par
Let us now consider the second relation (\ref{49e}) concerning $ L_5^{(c)(n)}(1)$. To isolate this quantity it is convenient to use the relations that generalize (B2)  and (B3) of ref. \cite{FN} :
\beq
\label{61e}
h_V^{(n)} = \xi^{(n)} + \epsilon_c \left( L_2^{(c)(n)} - L_5^{(c)(n)} \right) + O(\epsilon_b) + O(1/m_Q^2) \ ,
\eeq
\beq
\label{62e}
h_{A_1}^{(n)} = \xi^{(n)} + \epsilon_c \left( L_2^{(c)(n)} - {w-1 \over w+1}\ L_5^{(c)(n)} \right) + O(\epsilon_b) + O(1/m_Q^2)\ .
\eeq

We consider now the difference $h_V^{(n)} - h_{A_1}^{(n)}$, from which we can isolate the $\epsilon_c$ part by formally making $m_b \gg m_c$. Performing the calculation of the form factors in the BT model under the same conditions, and along the lines of \cite{DONG}, we find
\beq
\label{63e}
L_5^{(c)(n)} (1) = - \left( \Delta E^{(n)} + \overline{\Lambda} \right) {1 \over 2 \pi^2} \int p^2 dp\ \varphi^{(n)}(p)^* \varphi(p)
\eeq
$$ +\ {1 \over 6 \pi^2} \int p^2 dp\ p\sqrt{m^2+p^2} \left( \varphi^{(n)}(p)^*\varphi'(p) - \varphi^{(n)'}(p)^*\varphi(p) \right) \qquad (n \geq 0) .$$ 

\noi Taking into account the normalization/orthogonality condition we have
\beq
\label{64e}
L_5^{(c)(n)} (1) = - \left( \Delta E^{(n)} + \overline{\Lambda} \right) \delta_{n,0}
\eeq
$$ +\ {1 \over 6 \pi^2} \int p^2 dp\ p\sqrt{m^2+p^2} \left( \varphi^{(n)}(p)^*\varphi'(p) - \varphi^{(n)'}(p)^*\varphi(p) \right) \qquad (n \geq 0) .$$ 

\noi We obtain therefore, for the elastic case $n = 0$, since the second term cancels,
\beq
\label{65e}
L_5(1) = - \overline{\Lambda}\ .
\eeq

\noi that we found already in the BT model in \cite{DONG}, in agreement with the HQET result at zero recoil (\ref{54e}).\par
However, for the inelastic radially excited case, $n > 0$, the first term in (\ref{64e}) vanishes, and we are left with the expression
\beq
\label{66e}
L_5^{(c)(n)} (1) = {1 \over 6 \pi^2} \int p^2 dp\ p\sqrt{m^2+p^2} \left( \varphi^{(n)}(p)^*\varphi'(p) - \varphi^{(n)'}(p)^*\varphi(p) \right) \qquad (n > 0)
\eeq

\noi which is independent of $\Delta E$ and is therefore inconsistent with the relation (\ref{49e}) that we have obtained within HQET.\par
In conclusion, we find for the transitions to radial excitations ${1 \over 2}^- (n = 0) \to {1 \over 2}^- (n \not = 0)$ the same phenomenon that we found in \cite{DONG} for $1/m_Q$ inelastic form factors ${1 \over 2}^- \to {1 \over 2}^+$, ${1 \over 2}^- \to {3 \over 2}^+$ at zero recoil that depend on level spacings and on leading IW functions within HQET. 

\section{Conclusion}

Following the original work by Falk and Neubert for the elastic case, we have studied the $1/m_Q$ corrections of form factors for semi-leptonic transitions between $B$ mesons and radially excited $D$ mesons. In particular, we have formulated a generalization of Luke's theorem in this inelastic case and some relations that involve the derivative of the inelastic Isgur-Wise function at zero recoil $\xi^{(n)'}(1)\ (n > 0)$. We have proposed some methods to isolate this quantity, that could thus be determined from physical form factors. On the other hand, we emphasize the successes and failures of the relativistic Bakamjian-Thomas relativistic quark model concerning these corrections.

\par 

\end{document}